\documentclass[conference]{IEEEtran}
\usepackage{cite}
\usepackage{amsmath,amssymb,amsfonts}
\usepackage{algorithmic}
\usepackage{graphicx}
\usepackage{textcomp}
\usepackage{xcolor}
\usepackage{soul}
\usepackage[hyphens]{url}
\usepackage{hyperref}

\def\BibTeX{{\rm B\kern-.05em{\sc i\kern-.025em b}\kern-.08em
    T\kern-.1667em\lower.7ex\hbox{E}\kern-.125emX}}

\title{Retrospective: EIE: Efficient Inference Engine on Sparse and Compressed Neural Network}

\author{
    Song Han$^{1,3}$,
	Xingyu Liu$^{4}$,
	Huizi Mao$^{3}$,
	Jing Pu$^{5}$,
	Ardavan Pedram$^{2,6}$,
	Mark A. Horowitz$^{2}$,
	William J. Dally$^{2,3}$,
\\
$^1$MIT \quad $^2$Stanford \quad $^3$ NVIDIA \quad $^4$ CMU \quad $^5$ Google \quad $^6$ Samsung
}

\begin{document}
\vspace{-10pt}
\maketitle
\thispagestyle{plain}
\pagestyle{plain}



\begin{abstract}

EIE proposed to accelerate pruned and compressed neural networks, exploiting weight sparsity, activation sparsity, and 4-bit weight-sharing in neural network accelerators. Since published in ISCA'16, it opened a new design space to accelerate pruned and sparse neural networks and spawned many algorithm-hardware co-designs for model compression and acceleration, both in academia and commercial AI chips. In retrospect, we review the background of this project, summarize the pros and cons, and discuss new opportunities where pruning, sparsity, and low-precision can accelerate emerging deep learning workloads.

\end{abstract}

\section{What we did well}
We started this project as deep learning accelerators are bottlenecked by the memory footprint. Computation is cheap and memory is expensive. Existing algorithm and hardware stack accelerate the inference of a neural network ``as is.'' We asked, can we compress the model first? and we developed the ``Deep Compression'' \cite{han2015learning, han2015deep} technique that can compress the weights of a neural network by an order of magnitude by pruning and quantization. Since pruned weights become zero, and zero multiplied by anything is still zero, we can potentially save the computation and memory. However, the resulting neural network is sparse and irregular, which conflicts with massively parallel computing, and runs inefficiently on general-purpose hardware. 

EIE demonstrated that special-purpose hardware can make it cost-effective to do sparse operations with matrices that are upto 50\% dense - while in software, density must be much less than 1\% to overcome the overhead of the sparse package. 

EIE exploits both weight sparsity and activation sparsity. It stores the weights in compressed sparse column format, parallelizes the computation by interleaving matrix rows over the processing elements, and detects the leading non-zero in activations. It not only saves energy by skipping zero weights but also saves the cycle by not computing it. EIE supports fine-grained sparsity, and allows pruning to achieve a higher pruning ratio.

EIE adopted aggressive weight quantization (4bit) to save memory footprint. To maintain accuracy, EIE decodes the weight to 16bit and uses 16bit arithmetic. This W4A16 approach (4-bit weight, 16-bit activation) is different from the conventional W8A8 approach. Such a design has been reborn in large language models (LLM). The single batch text generation of these models is dominated by matrix-vector multiplication — same as EIE. It is memory-bounded, and the weight memory is the bottleneck, not the activation — 4bit weight and 16bit activation become attractive to save memory and maintain accuracy at the same time, as adopted by many software LLM inference engines.\footnote{4bit LLM projects such as: \href{https://arxiv.org/pdf/2210.17323.pdf}{GPTQ}, \href{https://arxiv.org/pdf/2306.00978.pdf}{AWQ}, 
\href{https://github.com/ggerganov/llama.cpp}{llama.cpp}, \href{https://github.com/mlc-ai/mlc-llm}{MLC LLM}} However, these software solutions use linear integer weights, rather than a Kmeans codebook to make the weight decoding simpler and the arithmetic cheaper.  

\begin{figure}[t]
        \centering
        \vspace{-5pt}
        \includegraphics[width=0.8\linewidth]{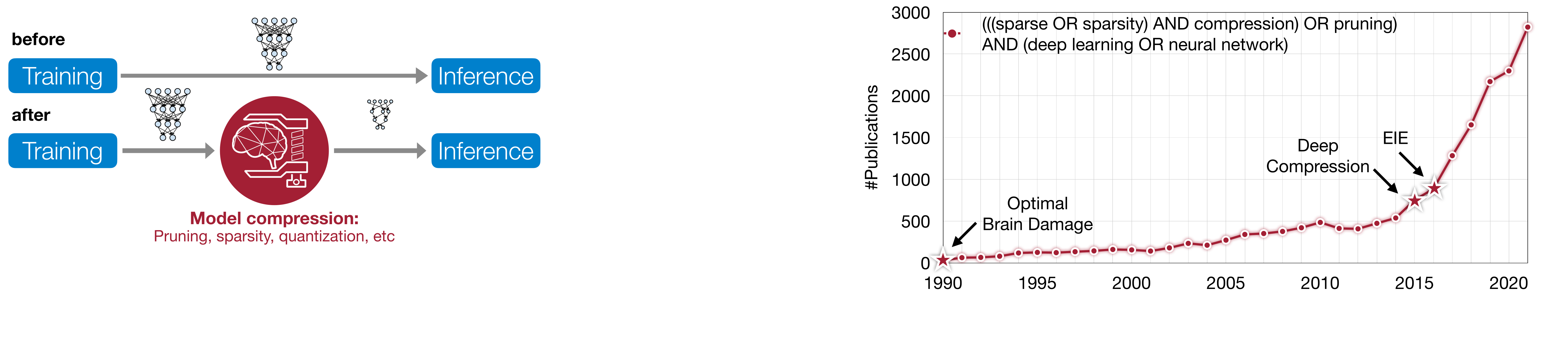}
        \vspace{-10pt}
        \caption{EIE opened a new opportunity to build hardware accelerator for sparse and compressed neural networks.}
        \vspace{-10pt}
        \label{fig:figure1}
\end{figure}

EIE demonstrates the opportunity for accelerator and neural network co-design. There’s plenty of room at the top to compress the neural network before accelerating it (Figure \ref{fig:figure1}). Deep Compression and EIE show the benefit of refactoring the design stack.

\section{Later work}
EIE generated a new wave of AI accelerator design by opening a new dimension: sparsity. 
 Cambricon-X \cite{zhang2016cambricon} proposes a prefix-sum-based indexing module and supports sparse CNNs.
SCNN~\cite{parashar2017scnn} utilizes outer product and scatter-add to process sparse CNN while maximizing the input data reuse. 
Pragmatic~\cite{albericio2017bit} skips bit-level zeros and eliminates ineffectual computations. 
UCNN~\cite{hegde2018ucnn} generalizes the sparsity problem to the repetition of weights with any value instead of zero.
Eyeriss V2~\cite{chen2019eyeriss} proposes a flexible interconnect and PE architecture to accelerate sparse CNN.
ExTensor~\cite{hegde2019extensor} hierarchically eliminates the computation in sparse tensor computations using an efficient intersection architecture.
SIGMA~\cite{qin2020sigma} proposes flexible interconnect to perform the distribution/reduction of sparse data for DNN training.
The Sparse Abstract Machine\cite{hsu2023sparse} targets sparse tensor algebra to reconfigurable and fixed-function spatial dataflow accelerators. 

EIE had substantial impacts on commercial AI chip design, leveraging pruning and sparsity for higher efficiency. NVDLA \cite{zhou2018research} gates the pruned weights to save energy. NVIDIA Sparse Tensor Core \cite{mishra2021accelerating} adopt structured 2:4 sparsity to speed up pruned models. Samsung NPU \cite{jang2021sparsity} uses a priority-based search algorithm to
skip zeros in activations. Ambarella CV22 \cite{ambarella} supports both structured and unstructured weight sparsity.



\section{Lessons}

Notwithstanding that EIE started sparse acceleration, this technique isn't as easily applied to arrays of vector processors. There are several improved designs that solved the issue, including Sparse Tensor Core \cite{zhu2019sparse,mishra2021accelerating} that adopted structured sparsity (N:M sparsity), where one PE becomes more effective PEs in a regular manner. Another improvement is load-balance-aware pruning \cite{han2017ese} to avoid PE starvation.

While EIE's special-purpose hardware is orders of magnitude more efficient than a software implementation of sparse M $\times$ V, the overhead of traversing the CSC structure is non-zero. One PE performs only one MAC, but is associated with many overhead structures, including pointer read, sparse matrix access, leading non-zero detector, etc. In EIE, the weight and index are both 4bit giving a 50\% storage overhead.   Other designs use structured sparsity or coarse-grained block sparsity to reduce storage and control overhead. 

EIE only accelerates fully connected layers. Later, SCNN\cite{parashar2017scnn}, Cambricon-X \cite{zhang2016cambricon} and Eyeriss-V2\cite{chen2019eyeriss} can also accelerate sparse convolution layers. EIE stores all the weights in SRAM. Commercially, Cerebras tried this path to put everything in SRAM. This setting is perfect for vision models, but not easy for LLM: the number of parameters of recent LLMs ranges from 10 billion to 100 billion, making it difficult to fit SRAM.

\begin{figure}[t]
        \centering
        \vspace{-5pt}
        \includegraphics[width=1\linewidth]{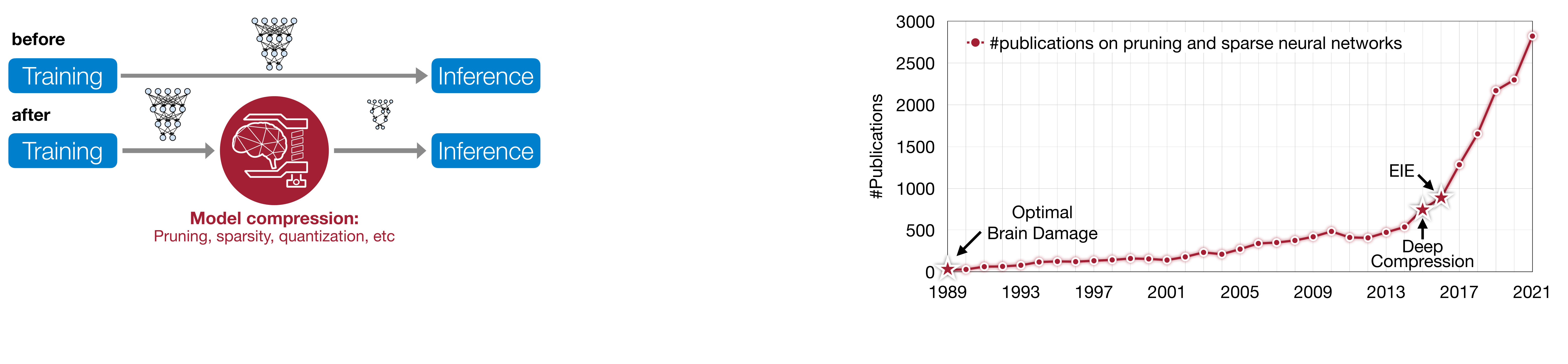}
        \vspace{-15pt}
        \caption{The number of publications on neural network pruning and sparsity quickly increased since 2015, including both algorithms and systems. \href{https://github.com/mit-han-lab/pruning-sparsity-publications}{Source}.}
         \vspace{-10pt}
        \label{fig:figure3}
\end{figure}

\section{New opportunities}
DNN architecture has witnessed rapid change. After EIE, we developed hardware-aware neural architecture search (NAS) techniques, ProxylessNAS \cite{cai2019proxylessnas} and Once-for-all \cite{cai2020once} that design small and fast models before model compression. 

The first principle of efficient AI computing is to be lazy: avoid  redundant computation, quickly reject the work, or delay the work. We show a few more examples.

After compressing the weights, the activation becomes the bottleneck. Therefore, we developed the MCUNet family\cite{lin2020mcunet, lin2021mcunetv2} that aggressively shrinks the activation for TinyML. MCUNet performs not only ImageNet classification but also detection with only 256KB SRAM and 1MB Flash on a microcontroller. By sparse update and low precision, we can even do on-device training under 256KB memory\cite{lin2022ondevice_mcunetv3}.

Generative AI: spatial sparsity persists in image editing or image in-painting; users don’t edit the whole image. So rather than generating the full image, sparsely generating where is edited \cite{li2022sige} can speed up inference. 

Transformer is a major neural architecture after EIE, and FC layer is back again. The attention layer has no weights to prune. However, not all tokens are useful: SpAtten \cite{wang2021spatten} proposes cascade token-pruning and gradually removes redundant tokens with the smallest attention score. It exploits ``progressive quantization'' that lazily fetches MSBs only, run inference; if the confidence is low, it fetches LSBs. 


Temporal sparsity exists in videos. Adjacent frames are similar. Rather than using expensive 3D convolution, temporal shift \cite{lin2019tsm} can efficiently exploit temporal redundancy with zero FLOPs.
Point cloud is spatially sparse. TorchSparse \cite{tang2022torchsparse} adaptively groups sparse matrices to trade computation for regularity. PointAcc \cite{lin2021pointacc} employs a sorting array to perform  sparse input-output mapping and avoid zero computation.

We envision future AI models will be sparse at various granularity and structures. 
Co-designed with specialized accelerators, sparse models will become more efficient and accessible.

\vspace{-7pt}
\small
\section*{Acknowledgements}
We thank Zhekai Zhang and Yujun Lin for the discussions and collecting data for the figure.
\vspace{-3pt}

\bibliographystyle{ieeetr}
\bibliography{refs}

\end{document}